\documentclass[12pt]{article}
\usepackage{amsmath,amssymb,amsfonts,graphicx,slashed,color,cite,bbm,dsfont}%, epsf, dsfont, dcolumn,yfonts}
\usepackage{authblk}
\usepackage[colorlinks=true,linkcolor=blue, citecolor=blue, linktoc=all]{hyper ref}

\DeclareMathAlphabet{\mathbbold}{U}{bbold}{m}{n}

\parskip=\baselineskip

\newcommand{\cDsl}{{{\cal D}\kern-.65em /\,}}
\newcommand{\cDslsm}{{{\cal D}\kern-.5em /\,}}
\newcommand{\nabslsm}{\nabla\kern-.55em /}
\newcommand{\pasl}{\pa\kern-.55em /}
\newcommand{\psl}{p\kern-.45em /}
\newcommand{\Dsl}{D\kern-.65em /}
\newcommand{\Asl}{A\kern-.55em /}
\newcommand{\nabsl}{\nabla\kern-.65em /\kern+.2em}
\newcommand{\qsl}{q\kern-.5em /}
\newcommand{\ksl}{k\kern-.5em /}
\newcommand{\rsl}{r\kern-.5em /}

\newcommand{\cDslLCsq}{{\stackrel{\circ}{\cDsl^{\kern2pt 2}}}}

\newcommand\cc[1]{#1^{^{\kern-6pt \circ}}\kern2pt}

\font\mybb=msbm10 at 11pt
\def\bb#1{\hbox{\mybb#1}}

\def\bC {\bb{C}}

\renewcommand{\a}{\alpha}
\renewcommand{\b}{\beta}

\newcommand{\ep}{\varepsilon}

\newcommand{\pa}{\partial}

\newcommand{\beq}{\begin{equation}}
\newcommand{\eeq}{\end{equation}}
\newcommand{\beqn}{\begin{eqnarray}}
\newcommand{\eeqn}{\end{eqnarray}}

\def\dalemb#1#2{{\vbox{\hrule height .#2pt
\hbox{\vrule width.#2pt height#1pt \kern#1pt
\vrule width.#2pt}
\hrule height.#2pt}}}

\newcommand{\cH}{\mathcal{H}}
\newcommand{\cS}{\mathcal{S}}

\newcommand{\bs}{\boldsymbol}
\newcommand{\cA}{\mathcal{A}}
\newcommand{\vx}{\vec{x}}
\newcommand{\vX}{\vec{X}}
\newcommand{\vn}{\vec{n}}
\newcommand{\vp}{\vec{p}}

\newcommand{\bd}{\boldsymbol{\delta}}

\newcommand{\cM}{\mathcal{M}}
\newcommand{\ba}{\mathbf{a}}
\newcommand{\cR}{\mathcal{R}}
\newcommand{\co}{\mathrm{open}}
\begin{document}

%\begin{center}
%\title
%\end{center}
%\centerline{[Draft Version of \today]}
%\vskip 2 cm
%\centerline{{\bf Good people}}

%
%\centerline{\it \uiucaddress}

%\centerline{\it \upennaddress}
%\vspace{2cm}
\title{Comments on Entanglement Entropy in String Theory}
\vspace{2cm}
\author[1,2]{Vijay Balasubramanian}
\author[1]{Onkar Parrikar}
\vspace{1cm}
\affil[1] {\small{David Rittenhouse Laboratory, University of Pennsylvania, 209 S.33rd Street, Philadelphia PA, 19104, U.S.A.}}
\affil[2]{\small{Theoretische Natuurkunde, Vrije Universiteit Brussel (VUB), and
International Solvay Institutes, Pleinlaan 2, B-1050 Brussels, Belgium.}}
\maketitle
\begin{abstract}
Entanglement entropy for spatial subregions is difficult to define in string theory because of the extended nature of strings. Here we propose a definition for Bosonic open strings using the framework of string field theory. The key difference (compared to ordinary quantum field theory) is that the subregion is chosen inside a Cauchy surface in the ``space of open string configurations''. We first present a simple calculation of this entanglement entropy in free light-cone string field theory, ignoring subtleties related to the factorization of the Hilbert space. We reproduce the answer expected from an effective field theory point of view, namely a sum over the one-loop entanglement entropies corresponding to all the particle-excitations of the string, and further show that the full string theory regulates ultraviolet divergences in the entanglement entropy. We then revisit the question of factorization of the Hilbert space by analyzing the covariant phase-space associated with a subregion in Witten's covariant string field theory. We show that the pure gauge (i.e., BRST exact) modes in the string field become dynamical at the entanglement cut. Thus, a proper definition of the entropy must involve an extended Hilbert space, with new stringy edge modes localized at the entanglement cut. 
\end{abstract}

%\pagebreak
\section{Introduction }
In this paper, we consider the following question: how does one define entanglement entropy for spatial subregions of the target space in string theory? In ordinary quantum field theory, the entanglement entropy of a subregion inside a spatial slice is defined as the von Neumann entropy of the reduced density matrix on the subregion. However, this definition exploits the inherent \emph{locality} in the Hilbert space of a ``standard'' quantum field theory, namely the fact that we can (at least in some lattice approximation) associate independent physical degrees of freedom to different points on a given Cauchy surface in spacetime.\footnote{More precisely, by locality here we mean that the total Hilbert space can be expressed as a tensor product $\mathcal{H}_{\Sigma} = \otimes_{\mathbf{x} \in \Sigma} \mathcal{H}_{\mathbf{x}}$ of Hilbert spaces associated with individual points $\mathbf{x}$ in a Cauchy surface $\Sigma$.} On the other hand, string theory describes fundamentally extended objects, and thus spacetime locality in the sense described above cannot be expected (except in an effective, low-energy sense). This tension between the non-locality of strings and the requirement of locality for a conventional definition of entanglement entropy for spatial subregions is the fundamental obstacle which forbids a straightforward definition.  Previous attempts at computing entanglement entropy in string theory have focussed on using the replica trick as a prescription \cite{Susskind:1994sm, Dabholkar:1994ai,Dabholkar:1994gg, He:2014gva, Mertens:2015adr, Mertens:2016tqv}, and the resulting quantity is often referred to as the conical entropy. This is not entirely satisfactory because it involves studying the perturbative string partition function on an off-shell background involving a conical singularity at the entanglement cut. Nevertheless, these calculations have yielded interesting insights; for instance, \cite{Susskind:1994sm} argued in the context of black hole entropy in closed string theory that the sphere diagram straddling the horizon gives the ``classical'' Bekenstein-Hawking entropy. The authors further interpreted this diagram as counting the states of open strings ending on the horizon, thus leading to a microscopic picture for the classical black hole entropy in terms of open strings ending on the horizon. This picture was recently explored further in the context of the string theory dual to large $N$ 2d Yang-Mills in \cite{Donnelly:2016jet}. In a separate recent development, \cite{He:2014gva, Mertens:2015adr, Mertens:2016tqv} computed the one-loop entanglement entropy across a Rindler horizon and argued for the UV finiteness of the entropy in closed superstring theory: 
\beq
S_{EE}  = s \frac{A_{\perp} }{\alpha'^4} + \cdots,
\eeq
where $s$ is a dimensionless, $O(1)$ constant. However, it would be nice to have an independent definition of entanglement entropy in string theory not relying on the replica trick. 

Before moving to strings, let us briefly revisit the setup for studying entanglement in conventional theories of point particles. In this case, it is convenient to use the quantum field theory description instead of the ``first-quantized'' wordline description. That is, instead of using the worldline variables (the position and intrinsic metric $X^{\mu}(\tau), g_{\tau\tau}(\tau)$), one switches over the to the quantum field operator 
$$\Phi(X^{\mu}),$$ 
which is an operator-valued function on the target spacetime $M$. One then picks a Cauchy surface $\Sigma$ in $M$, which for simplicity we can choose to be $X^0 = 0$. Since operators separated along $\Sigma$ commute, one can think of the operators at each point on $\Sigma$ as independent. In studying the spatial entanglement structure of a  state, one then partitions $\Sigma$ into two (or more) regions, and computes the reduced density matrix on a given subregion, which corresponds to tracing out all the operators outside the region of interest.

Now we wish to follow the same line of thought for string theory. For simplicity, we will consider bsosonic open string theory on Minkowski spacetime $M=\mathbb{R}^{1,25}$, but we expect that our setup can straightforwardly be generalized to superstrings, and even to closed strings at some level. Following the logic outlined above, we wish to pass from the worldsheet description to a ``field theory'' description. Naturally, since string theory describes strings as opposed to point particles, the corresponding \emph{string field operator} $\Phi$ is an operator-valued function not on the target spacetime $M$, but on ``\emph{the space of all open strings in spacetime}''
$$\Phi[X^{\mu}(\sigma)], $$
where as mentioned above, $X^{\mu} : [0, \pi] \to M$ is an open string configuration in $M$. We will call this ``space of all open strings in spacetime'' $\cM_{\co}$. (If we were considering closed string theory, then $\cM_{\mathrm{closed}}$ would be the loop space of $M$.) More precisely, the string fields also depend on ghost-configurations, but since we will not be interested in entanglement cuts along anti-commuting directions, we suppress them for now. We can be a little more explicit about $\cM_{\co}$ by mode-expanding the open-string configuration
\beq
X^{\mu}(\sigma) = X_0^{\mu} + \sum_{n=1}^{\infty} 2X_{n}^{\mu} \cos(n\sigma),
\eeq 
and treating the coefficients $X^{\mu}_n$ as coordinates on $\cM_\co$. The next step is to pick a Cauchy surface $\cS$ inside $\cM_{\co}$. We can choose $\cS$, for instance, to be the surface consisting of open strings with the time component of the center of mass coordinate $X^0_0$ fixed, or somewhat more generally  $t_{\mu}X_0^{\mu} $ fixed, where $t_{\mu}$ is a time-like vector.\footnote{Another natural choice in Witten's cubic string field theory \cite{Witten:1985cc} is to consider the space of all open strings with the midpoint time coordinate fixed, i.e. $t_{\mu}X^{\mu}(\pi/2) = 0$. More on this in later sections.} The phase space of the theory can then be established by constructing a symplectic form on $\cS$ \cite{Witten:1986qs}. From standard reasoning, string-fields at different points on $\cS$ commute
\beq
\Big[\Phi[X^{\mu}(\sigma)], \Phi[Y^{\mu}(\sigma)] \Big] = 0, \;\;\;\; X^{\mu},\,Y^{\mu }\in \cS.
\eeq
\begin{figure}[t]
\centering
\includegraphics[height=5cm]{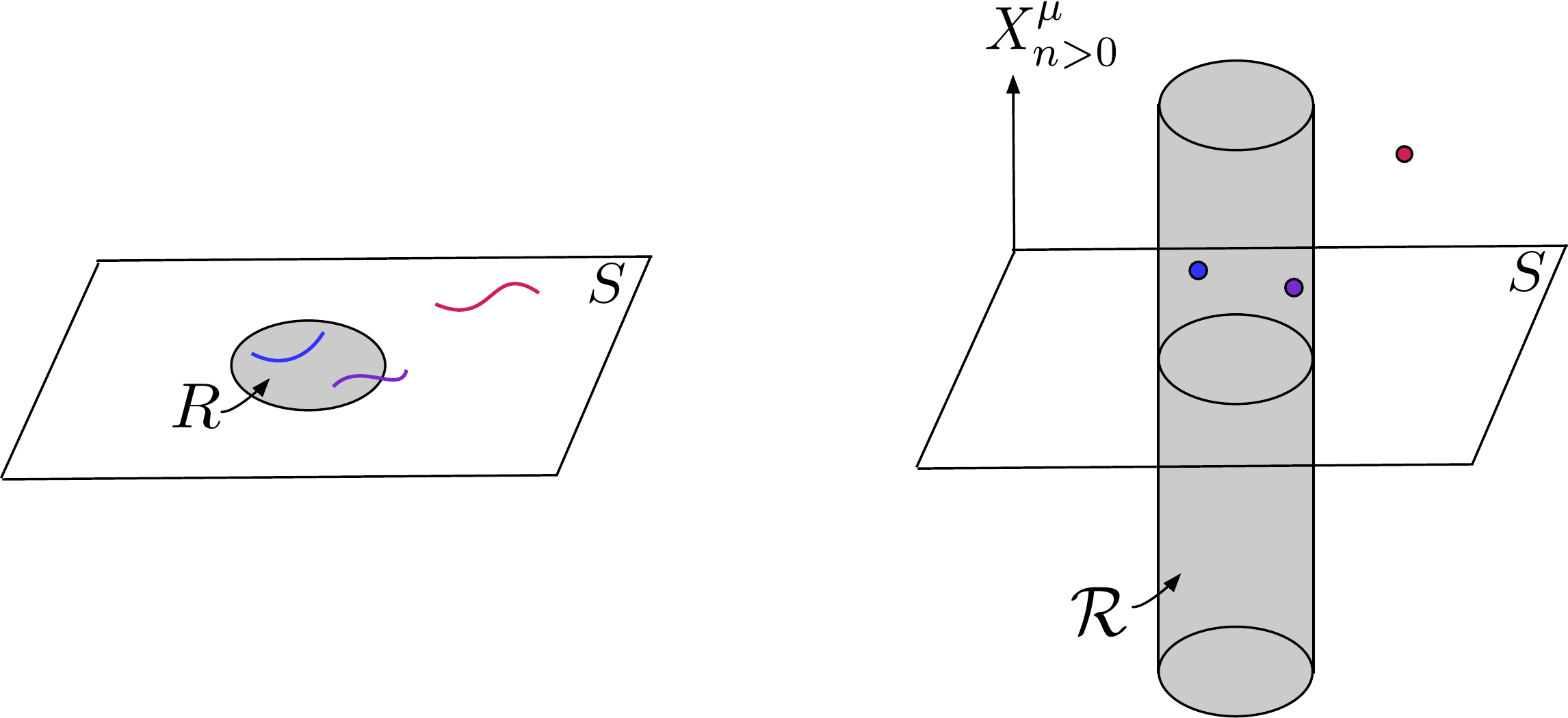}
\caption{\small{(Left) From a spacetime point of view, we're interested in the degrees of freedom inside a subregion $R$ of the Cauchy Surface $S$. However, in string theory we can have strings which either lie in $R$ (blue), or straddle the cut $\pa R$ (violet), or lie outside $R$ (red). (Right) In string field theory, we get around this issue by considering a subregion $\cR$ inside the space of open string configurations. One natural choice of the subregion is the trivial extension of $R$ along the stringy directions $X^{\mu}_{n>0}$; the open string configurations from the left panel are displayed as points. }}
\end{figure}
Equivalently, we can think of string field operators at different points on $\cS$ as being independent degrees of freedom. In order to study the entanglement properties of the vacuum in this theory, we can partition the surface $\cS$ into subregions. Note that this is very different from the situation in ordinary field theory -- here we are required to partition not a Cauchy surface $\Sigma$ in the target spacetime $M$, but instead an infinite-dimensional surface $\cS$ in the space of open string configurations $\cM_\co$. A partition of $\Sigma$ of course does not extend to a partition of $\cS$ in a unique way -- indeed given a subregion $R \subset \Sigma$, there are many ways to extend this to a subregion $\cR \subset \cS$, and the results of our computations will depend on the choice of extension. One natural choice from the spacetime point of view is to extend the subregion as follows (see figure 1):
\beq \label{ext}
\cR = \left\{ X^{\mu}(\sigma) \in \cS \; | X^{\mu}_0 \in R \right\}.
\eeq
$\cR$ consists of all strings with their center-of-mass inside the spacetime region $R$; this does \emph{not} mean that the entire string is contained in $R$. In particular, $\cR$ includes strings which straddle the entanglement cut $\pa R$ from a spacetime point of view. (Of course, another interesting possibility is to choose $\cR$ such that the entire string is contained inside $R$, but we will not attempt this here; we will briefly return to this point in the Discussion.) In any case, once we choose the subregion $\cR$, the reduced density matrix over $\cR$, and subsequently the entanglement entropy, can be defined and computed in the standard way, much like  in conventional quantum field theory. So string field theory provides a potentially feasible and well-defined way to study the entanglement structure of the vacuum in string theory. In this paper, we wish to take the first steps in this direction. 

The rest of the paper is organized as follows. In section \ref{LCG}, we will compute the entanglement entropy (following the definition explained above) in free string field theory ($g_s \to 0$) in the light-cone gauge.  In this computation we will ignore subtleties associated with the factorization of the Hilbert space. This will serve mainly as a sanity check on our proposed definition, as we will show that the entanglement entropy agrees with the expected answer from a spacetime effective field theory point of view. Then, in section \ref{CPS}, we will recall that string field theory has a gauge symmetry which interferes with factorization of the Hilbert space on $\cS \subset \cM_{\co}$.  We will address this issue by  analyzing the covariant phase space corresponding to a subregion in covariant string field theory (again in the free limit), following the framework of Donnelly and Friedel \cite{Donnelly:2016auv}.  We will show that BRST exact modes in string field theory become dynamical at the entanglement cut, much like in conventional gauge field theories where  edge modes appear in the computation of entanglement entropy.

\section{Entanglement entropy in Light-cone gauge} \label{LCG}
In this section, we will consider \emph{free} bsosonic, open string field theory (OSFT) in $D=26$ spacetime dimensions. We will use the light-cone gauge for OSFT following \cite{Kaku:1974zz,Kaku:1974xu,THORN19891, Martinec:1993jq, Lowe:1993ps} in our analysis; see also \cite{Hata:1995di} which studies covariant string field theory in Rindler space.\footnote{The approach in \cite{Hata:1995di} is similar in spirit to our calculations in section 2. We thank Thomas Mertens for bringing \cite{Hata:1995di} to our attention, after the first version of our paper was uploaded to the arXiv.} We will consider a half-space $R$ in spacetime, and the corresponding extension $\cR$ (as in equation \eqref{ext}) to the space of all open strings.  The computation in this section is a simplified version of the full story, because it ignores the lack of factorization of the OSFT Hilbert Space, an issue that we will address in the next section. 
%For instance, we do not dwell here on whether the the Hilbert space of OSFT actually factorizes, i.e., $\cH_{\cS} \stackrel{?}{=} \cH_{\cR} \otimes \cH_{\bar{\cR}}$. We will come back to these subtleties in the next section, where we will carefully analyze the covariant phase space of string theory over subregions, issues related to gauge-invariance etc.

%We begin reviewing some basic definitions, following the above authors. 
We choose light-cone coordinates:
\beq
X^{\pm} = X^0 \pm X^{D-1}
\eeq
in terms of which the metric of Minkowski spacetime is
\beq
ds^2 = -dX^+dX^- + d\vec{X}^id\vec{X}^i. 
\eeq
In the light cone gauge, we fix $X^+$ to be $\sigma$-independent:
\beq
X^+(\sigma) = x^+ 
\eeq
\beq
X^-(\sigma) = x^-_0 + \cdots
\eeq
\beq
\vec{X}^i (\sigma) = x^i_0  + \sum_{\ell=1}^{\infty} 2 \, x^i_{\ell} \, \cos(\ell \sigma).
\eeq
where the $\cdots$ denote oscillator modes along the $-$ direction, which get fixed in terms of the remaining modes by the constraint that the worldsheet stress tensor vanishes. The string field takes the form $\Phi[x^+,x^-_0, \vec{X}^i(\sigma)]$, with an equation of motion
\beq
-2\frac{\pa^2 \Phi}{\pa x^+ \pa x_0^-} =  \int_0^{\pi} d\sigma\left( - \frac{\delta^2}{\delta \vec{X}^2(\sigma)}  + \pa_{\sigma}\vec{X}^2(\sigma)\right) \Phi .
\eeq
where we have set $\alpha' = 1$. We can write equivalently this in terms of the following action 
\beq
S_0 =  \frac{1}{2}\int dx^+dx^-_0\int [d\vec{X}(\sigma)]  \left\{2\Phi  \frac{\pa^2}{\pa x^+ \pa x_0^-} \Phi+ \Phi\int_0^{\pi} d\sigma\left( - \frac{\delta^2}{\delta \vec{X}^2(\sigma)}  + \pa_{\sigma}\vec{X}^2(\sigma)\right)\Phi \right\}. 
\eeq

It seems natural to quantize this system on the $x^+= 0$ surface, i.e. by treating $x^+$ as time. However, light-front quantization is fairly subtle, because if we want to regard $x^+$ as time, then the Lagrangian is first order in $\pa_+\Phi$ \cite{Burkardt:1995ct}. Consequently, it is not consistent to think of $\Phi$ and $\Pi \sim \pa_-\Phi$ as independent operators. In other words, light-front quantization amounts to quantizing a constrained system, and one has to use Dirac brackets. In order to avoid these complications, we will follow a different approach and define new coordinates
\beq
x^- \rightarrow x^-,\qquad x^+ \rightarrow x^+ + \frac{1}{2}\ep x^-
\eeq
In terms of these coordinates, the action becomes 
%\beq
%-\left(2\frac{\pa^2 }{\pa x^+ \pa x_0^-}+\ep \frac{\pa^2 }{\pa x^+ \pa x^+}\right)\Phi  = \alpha' \int_0^{\pi} d\sigma\left( - \frac{\delta^2}{\delta \vec{X}^2(\sigma)}  + \pa_{\sigma}\vec{X}^2(\sigma)\right) \Phi 
%\eeq
%which comes from the Lagrangian
\beq
\begin{split}
S_0 = \frac{1}{2} \int dx^+dx^-_0\int [d\vec{X}(\sigma)]  &\left\{ \Phi  \left(2\frac{\pa^2 }{\pa x^+ \pa x_0^-}+\ep \frac{\pa^2 }{\pa x^+ \pa x^+}\right) \Phi \right.\\
& + \left. \;\Phi\int_0^{\pi} d\sigma\left( - \frac{\delta^2}{\delta \vec{X}^2(\sigma)}  + \pa_{\sigma}\vec{X}^2(\sigma)\right)\Phi \right\} .
\end{split}
\eeq
Now the action is quadratic in $\pa_+\Phi$ (after integration by parts), and we can quantize this system canonically.  The conjugate momentum is
\beq
\Pi = \pa_-\Phi + \ep \pa_+ \Phi.
\eeq

We can expand the string field operator in terms of solutions to the equation of motion derived from the above action.  This gives 
\beq
\begin{split}
\Phi[x^+,x^-_0, \vX(\sigma)] &= \int \frac{d^{D-2}\vp}{(2\pi)^{D-2}} \int_{-\infty}^{\infty} \frac{dp^+}{2\pi}\frac{1}{\sqrt{2(p^++\ep p^-)}} \\
&\times \sum_{\{\vn_{\ell}\}} \left(\ba_{p^+, \vp, \{\vn_{\ell}\}} e^{-i(p^+ x_0^- +p^-x^+-\vp\cdot \vx_0)}\prod_{\ell=1}^{\infty} f_{\vn_{\ell}}(\vx_{\ell}) + h.c.\right),
\end{split}
\eeq
where  in the oscillator directions we have the simple-harmonic oscillator wavefunctions
\beq
f_{\vn_{\ell}}(\vx_{\ell}) = \prod_{i=1}^{D-2} H_{n^i_{\ell}}(x^i_{\ell}) e^{-\ell (x_{\ell}^i)^2},
\eeq
and there is a dispersion relation
\beq
p^-+\frac{p^+}{\ep} = \left(\frac{{p^+}^2}{\ep^2}+\frac{\vp^2+\sum_{\ell=1}^{\infty} \ell \sum_{i=1}^{D-2} n_{\ell}^i}{\ep}\right)^{1/2}.
%&=& \frac{|p^+| - p^+}{\ep} + \frac{\vp^2+m^2+\sum_{\ell=1}^{\infty} \ell \sum_{i=1}^{D-2} n_{\ell}^i}{2|p^+|} +O(\ep).
\eeq
Further, $\ba^{\dagger}_{p^+, \vp, \{\vn_{\ell}\}}$ and $\ba_{p^+, \vp, \{\vn_{\ell}\}}$ are operators which create and annihilate strings with the specified mode occupation numbers.\footnote{Not to be confused with the usual first-quantized string oscillators $\alpha^{\mu}_{n}$ which create modes on a given string.} They satisfy the commutation relations
\beq
\left[\ba_{p^+, \vp, \{\vn_{\ell}\}}, \ba^{\dagger}_{{p^+}', \vp', \{\vn'_{\ell}\}}\right] = (2\pi)^{D-2}\delta^{D-2}(\vp - \vp') 2\pi \delta(p^+ - {p^+}') \delta_{\{\vn_{\ell}\}, \{\vn'_{\ell}\}}.
\eeq 
The normalization for the oscillators is chosen such that
\beq
\left[ \Phi (x^+, x^-_0, \vec X(\sigma)), \Pi (x^+, y^-_0, \vec Y(\sigma')) \right] = -i\delta(x^-_0  - y^-_0) \delta(\vec{X}(\sigma) - \vec{Y}(\sigma')).
\eeq
The vacuum $|\mathbf{0}\rangle $ of string field theory satisfies
\beq
\ba_{p^+, \vp, \{\vn_{\ell}\}} | \mathbf{0} \rangle = 0.
\eeq
Of course, in bosonic string theory this vacuum is unstable as indicated by the presence of a tachyon in the spectrum, but we will continue to work with this state here. 

In order to compute the entanglement entropy, we will need the various equal-time two-point functions of the string field $\Phi$ and its momentum.  The $\Phi-\Phi$ correlator is
\beq
G_{\Phi,\Phi}(0,x^-_0, \vec X|0,y^-_0, \vec Y) =  \langle \mathbf{0} |  \Phi (0, x^-_0, \vec X(\sigma)) \Phi (0, y^-_0, \vec Y(\sigma')) | \mathbf{0} \rangle 
\eeq
is given by
%\beq
%G_{\Phi,\Phi}(0,x^-_0, \vec X|0,y^-_0, \vec Y) =  \int \frac{d^{D-2}\vp}{(2\pi)^{D-2}}  \int_{-\infty}^{\infty} \frac{dp^+}{2\pi}\sum_{\{\vn_{\ell}\}}\frac{1}{2(p^++\ep p^-)} \;e^{ip^+ (y_0^- - x_0^-) - i\vp\cdot (\vec{y}_0- \vec{x}_0)} \prod_{\ell=1}^{\infty} f_{\vn_{\ell}}(\vx_{\ell}) f^*_{\vn_{\ell}}(\vec{y}_{\ell}).
%\eeq
%where recall that 
%\beq
%p^+ + \ep p^- = \ep \left(\frac{{p^+}^2}{\ep^2}+\frac{\vp^2+m^2+\sum_{\ell=1}^{\infty} \ell \sum_{i=1}^{D-2} n_{\ell}^i}{\ep}\right)^{1/2}.
%\eeq 
%By rescaling $p^+ \to p^+/\sqrt{\ep}$, we can rewrite the above two-point function as
\beqn
G_{\Phi, \Phi}(0,x^-_0, \vec X|0,y^-_0, \vec Y) &=&  \int \frac{d^{D-2}\vp}{(2\pi)^{D-2}}  \int_{-\infty}^{\infty} \frac{dp^+}{2\pi}\sum_{\{\vn_{\ell}\}}\frac{1}{2\omega} e^{ip^+ \sqrt{\ep}(y_0^- - x_0^-) - i\vp\cdot (\vec{y}_0- \vec{x}_0)} \prod_{\ell=1}^{\infty} f_{\vn_{\ell}}(\vx_{\ell}) f^*_{\vn_{\ell}}(\vec{y}_{\ell})\nonumber\\
%&=&\int_0^{\infty} \frac{ds}{\sqrt{4\pi s}} \int \frac{d^{D-2}\vp\, dp^+}{(2\pi)^{D-1}} \sum_{\{\vn_{\ell}\}}e^{-s\omega^2} e^{ip^+ \sqrt{\ep}\delta x_0^- - i\vp\cdot \delta\vec{x}_0} \prod_{\ell=1}^{\infty} f_{\vn_{\ell}}(\vx_{\ell}) f^*_{\vn_{\ell}}(\vec{y}_{\ell}) 
\eeqn
where we have defined %$\delta x_0 = y_0 - x_0$, and
\beq
\omega =  \left({p^+}^2+\vp^2+\sum_{\ell=1}^{\infty} \ell \sum_{i=1}^{D-2} n_{\ell}^i\right)^{1/2}.
\eeq 
%We can carry out the $(p^+,\vp)$-integral
%\beq
%K^{(s)}_0(\delta x^-, \delta\vx) = \int \frac{d^{D-2}\vp\,dp^+}{(2\pi)^{D-1}}  e^{ip^+\sqrt{\ep}\delta x^-_0-i \vp \cdot \delta \vx_0 - s\vp^2 - s {p^+}^2} = \left( \frac{1}{4\pi s}\right)^{\frac{D-1}{2}}e^{-\frac{1}{4s} \left(\ep \delta {x_0^-}^2 + \delta \vx_0^2\right)}
%\eeq
%Additionally, we can carry out the sum over $\{\vec n_{\ell}\}$ by using the heat-kernel for the simple-harmonic oscillator
%\beqn
%K^{(s)}_{\ell}( \vx_{\ell}, \vec{y}_{\ell}) &=& \sum_{\vn_{\ell}} f_{\vn_{\ell}}(\vx_{\ell}) f^*_{\vn_{\ell}}(\vec{y}_{\ell}) e^{- \ell \frac{s}{\alpha'} \sum_{i=1}^{D-2} n_{\ell}^i}\\
%&=& e^{ \ell (d-2)\frac{s}{2\alpha'}}\langle \vec{x}_{\ell} | e^{-\ell \frac{s}{\alpha'} \sum_{i}\widehat{H}_{\ell}^i} |\vec{y}_{\ell}\rangle\nonumber\\
%&=& \left( \frac{2\ell}{\pi (1- e^{-\frac{4\ell s}{\alpha'}})}\right)^{\frac{d-2}{2}} \exp\left(\frac{\ell}{\sinh(\frac{2 \ell s}{\alpha'})}\left[(\vx_{\ell}^2+ \vec{y}_{\ell}^2)\cosh \left(\frac{2\ell s}{\alpha'}\right) - 2\vx_{\ell} \cdot \vec{y}_{\ell}\right]\right)\nonumber
%\eeqn
%So we find
%\beq
%G_{\Phi,\Phi}(x^-_0, \vec X|y^-_0, \vec Y) =  \int_0^{\infty} \frac{ds}{\sqrt{\pi s}}e^{-sm^2} K^{(s)}_0(\delta x^-, \delta\vx) \prod_{\ell=1}^{\infty} K_{\ell}^{(p^+, \delta x^+)}( \vx_{\ell}, \vec{y}_{\ell}) 
%\eeq
%It is easy to check that if we ignore all the string-oscillator modes, than this gives the right 2-point function for a massive scalar (the tachyon). 
Similarly, the other two-point functions are given by
\beq
G_{\Phi,\Pi}(0,x^-_0, \vec X|0,y^-_0, \vec Y) = -\frac{i}{2} \delta(x^-_0  - y^-_0) \delta(\vec{X}(\sigma) - \vec{Y}(\sigma')),
\eeq
\beq
G_{\Pi,\Pi}(0,x^-_0, \vec X|0,y^-_0, \vec Y) =- \int \frac{d^{D-2}\vp}{(2\pi)^{D-2}}  \int_{-\infty}^{\infty} \frac{dp^+}{2\pi}\sum_{\{\vn_{\ell}\}}\frac{\varepsilon\omega}{2} e^{ip^+ \sqrt{\ep}(y_0^- - x_0^-) - i\vp\cdot (\vec{y}_0- \vec{x}_0)} \prod_{\ell=1}^{\infty} f_{\vn_{\ell}}(\vx_{\ell}) f^*_{\vn_{\ell}}(\vec{y}_{\ell}).
\eeq

In order to compute the entanglement entropy,  we will use the algebraic definition in terms of correlation functions \cite{2003JPhA...36L.205P, Casini:2009sr}. Let us consider the half-space $x^- >0$ on the spacetime Cauchy surface $x^+=0$.\footnote{Our general conclusions do not depend on this particular choice of half-space (i.e., we could have equally well picked other regions), although of course the specific expression will do so.}  As discussed in the introduction, we wish to extend this to a subregion $\cR$ in the space of open strings. This extension is not unique, but we use the simplest one,
\beq
\cR = \left\{x_0^-, \vec{X}(\sigma) | x_0^- > 0\right\},
\eeq
which is the analog of Figure 1 for a half-space. The basic idea in the algebraic method \cite{2003JPhA...36L.205P, Casini:2009sr} is that for bosonic Gaussian systems, the reduced density matrix is the exponential of a bilocal operator of the form:
\beq
\rho \sim e^{-\sum_{ij} \left(  A_{ij} \Phi_i\Phi_j  +B_{ij} \Phi_i \Pi_j + C_{ij} \Pi_i\Pi_j \right)},
\eeq
where the indices $i,j...$ denote spatial coordinates restricted to the region $R$ for ordinary quantum fields on a lattice.  In the continuum the sums on indices become integrals, and in the present case they become integrals on the string coordinates restricted to $\cR$. By definition, the reduced density matrix should reproduce the correct correlation functions for operators inside $\cR$, so the matrices $A, B$ and $C$ can be recovered from the knowledge of two-point functions of $\Phi$ and $\Pi$ restricted to $\cR$. 

We will not go through the details of this calculation here, but we merely state the result. Let us define the matrix
\beq
C^2(x^-_0, \vec X(\sigma)|y^-_0, \vec Y(\sigma)) = -\int_0^{\infty} dz_0^-\int [d\vec{Z}(\sigma)] G_{\Phi,\Phi}(x^-_0, \vec X|z^-_0, \vec Z)G_{\Pi,\Pi}(z^-_0, \vec Z|y^-_0, \vec Y)
\eeq
where  the $(z_0^-, \vec{Z}(\sigma) )$ integral is restricted to the subregion $\cR$. In terms of the matrix $C$, the entanglement entropy is formally given by\footnote{In order to make these manipulations concrete (i.e., to make sense of the infinite-dimensional integrals), we can discretize the string and pick a lattice in the target space.}
\beq
S_{EE}(\cR)  = \mathrm{Tr}\Big( (C+1/2)\ln\,(C+1/2) - (C-1/2)\ln\,(C-1/2) \Big).
\eeq

Since we know the two-point functions of the string fields, we can easily compute $C^2$:
\beq
C^2(x^-_0, \vec X|y^-_0, \vec Y) =\sum_{\{\vn_{\ell}\}}\Gamma_{ \{\vn_{\ell}\}}  (x_0^-,\vec{x}_0; y_0^-, \vec{y}_0)\;\prod_{\ell=1}^{\infty} f_{\vn_{\ell}}(\vx_{\ell}) f^*_{\vn_{\ell}}(\vec{y}_{\ell}), \label{bd}
\eeq
where 
\beq
\Gamma_{ \{\vn_{\ell}\}}  = \frac{i\sqrt{\ep}}{4}\int \frac{d^{D-2}\vp}{(2\pi)^{D-2}}e^{i\vp \cdot(\vec{x}_0 - \vec{y}_0)}\int_{-\infty}^{\infty}\frac{dp^+}{2\pi}\int_{-\infty}^{\infty}\frac{dq^+}{2\pi} \frac{e^{-i\sqrt{\ep}(p^+x_0^- - q^+y_0^-)}}{p^+-q^+} \left(\frac{{q^+}^2++\vp^2+\mu_{\{\vn_{\ell}\}}^2}{{p^+}^2+\vp^2+\mu_{\{\vn_{\ell}\}}^2}\right)^{1/2},
\eeq
\beq
\mu_{\vp,\{\vn_{\ell}\}}^2 =   \sum_{\ell=1}^{\infty}\ell \sum_{i=1}^{D-2}n^i_{\ell}.
\eeq
The reader can check that $\Gamma$ is identical to the corresponding matrix $-G_{\phi\phi}\cdot G_{\pi\pi}$ for a real scalar field of mass $\mu^2$ (restricted to the half space).   

It is clear from equation \eqref{bd} that the matrix $C^2$ is block-diagonal in terms of the  oscillator excitations of the open string, namely $|0\rangle, \alpha^i_{-1} |0\rangle, \alpha^{i}_{-1}\alpha_{-1}^{j} | 0\rangle, \cdots $ ; namely Eq.~\ref{bd} has single sum over  $\vec{n}_\ell$. This is essentially a direct consequence of the fact that the subregion $\cR$ is a direct product of the spacetime subregion $R$ and the oscillator directions $X^{\mu}_{n>0}$. Therefore, the entanglement entropy takes the form
\beq \label{sum}
S_{EE}  = \sum_{\{\vn_{\ell}\}} S_{EE}(\mu_{\{\vn_{\ell}\}}^2)  ,
\eeq
where $S_{EE}(\mu_{\{\vn_{\ell}\}}^2)$ is the entanglement entropy for a scalar of mass $\mu_{\{\vn_{\ell}\}}$, as should be clear from the above interpretation of the structure of $\Gamma$. In other words, every excitation in the list $|0\rangle, \alpha^i_{-1} |0\rangle, \alpha^{i}_{-1}\alpha_{-1}^{j} | 0\rangle, \cdots $ adds one scalar degree of freedom worth of entropy, with the appropriate mass. This result makes sense physically, because in the free limit, the various oscillator modes of the string are entirely decoupled, and should contribute independently to the entanglement. So  we can sum the entanglement entropies corresponding to the tachyon, the photon, etc. where the tachyon contributes one scalar degree of freedom, the photon contributes 24 degrees of freedom etc. Note that since we're computing the entanglement entropy algebraically, we do not encounter the additional ``contact terms'' which appear in the conical entropy \cite{Kabat:1995eq, He:2014gva}; we will come back to this in the next section. 

In order to obtain an explicit expression for $S_{EE}$ we would ordinarily have to diagonalize the matrix $\Gamma_{\{\vn_{\ell}\}} $. However, this is unnecessary in the present case; the half-space entropy of a massive scalar field in $d$-dimensions is already well-known (see \cite{He:2014gva}, for instance)
\beq \label{sm}
S_{EE}(m^2) = \frac{1}{6}A_{\perp} \int_{\ep^2}^{\infty} \frac{ds}{2s}\frac{1}{(4\pi s)^{d/2-1}} e^{-sm^2},
\eeq
where $A_{\perp}$ is the area of the entanglement cut, and $\ep$ is an ultraviolet cut-off. This integral is UV divergent, and takes the form
\beq
S_{EE} (m^2)= \frac{1}{12} \frac{A_{\perp}}{(4\pi)^{n}n!}  \left(\frac{e^{-m^2\ep^2}}{\ep^{2n}} \sum_{k=0}^{n-1}(-1)^k(n-k-1)!(m\ep)^{2k}-(-1)^{n}\left(2 \ln(m\ep)+\gamma\right) + O(m\ep)\right)
\eeq
where $n = \frac{d}{2}-1$. However, string theory softens these divergences to a great extent. In order to see this, we first first perform the sum in equation \eqref{sum} before carrying out the $s$-integral (restoring $\alpha'$):
\beq
S_{EE}  = \frac{A_{\perp}}{6}  \int_{\ep^2}^{\infty} \frac{ds}{2s}\frac{1}{(4\pi s)^{d/2-1}} \sum_{N=0}^{\infty} \mathrm{deg}_N\,e^{-s\frac{(N-1)}{\alpha'}},
\eeq
where $\mathrm{deg}_N$ is the number of states at mass $m^2 = \frac{N-1}{\alpha'}$. Defining $q= e^{-s/\alpha'}$ and writing $N = \sum_{\mu, n} n N_{\mu,n}$, we can perform the sum in the standard way \cite{polchinski_1998}:
\beqn \label{exp}
 \sum_{N=0}^{\infty} \mathrm{deg}_N\,e^{-s\frac{(N-1)}{\alpha'}}  &=& q^{-1} \prod_{\mu=1}^{24}\prod_{n=1}^{\infty}\left( \sum_{N_{\mu,n}=0}^{\infty} q^{ n N_{\mu,n} }\right)\nonumber\\
 &=&\frac{1}{q} \left(\prod_{n=1}^{\infty}\frac{1}{(1-q^n)}\right)^{24}\nonumber\\
 &=& \eta(is/2\pi \alpha')^{-24},
 \eeqn
where the Dedekind eta function is defined as
\beq
\eta(\tau) = q^{1/24} \prod_{n=1}^{\infty}(1-q^n),\;\;\; q = e^{2\pi i \tau}.
\eeq
So we have the final result
\beq
S_{EE}  = \frac{1}{6}\frac{A_{\perp}}{(8\pi^2\a')^{d/2-1}}  \int_{\ep_0^2}^{\infty} \frac{dt}{2t}\frac{1}{t^{d/2-1}} \eta(it)^{-24},\;\;\;\ep_0^2 = \frac{\ep^2}{2\pi\a'}.
\eeq

Let us now analyze the divergences in the entanglement entropy. As is familiar from standard considerations in string theory, there are two types of divergences in the above integral: $t \to 0$ (UV divergence) and $t \to \infty$ (IR divergence). We can analyze the $t \to \infty $ divergence by expanding the eta function:
\beq
\eta(it)^{-24} = e^{2\pi t} + 24 + O(e^{-2\pi t})
\eeq
where the various terms this expansion should be thought of, similar to the left hand side of equation \eqref{exp}, as corresponding to on-shell states of different $m^2$. The first term (which is due to the tachyon) gives a divergence in the $t \to \infty$ limit, but this is unphysical anyway because the tachyon will be projected out in the superstring. The remaining terms do not give any divergences as $t\to \infty$. Now consider the $t \to 0$ (UV) limit. Here, we can switch to $s = 1/t$ and use the property
\beq
\eta(i/t) = t^{1/2} \eta(it),
\eeq 
to re-write the entropy as
\beq
S_{EE}  = \frac{1}{6}\frac{A_{\perp}}{(8\pi^2\a')^{d/2-1}}  \int_0^{1/\ep_0^2} \frac{ds}{2s}\eta(is)^{-24}.
\eeq
Now we again expand in the large $s$ limit using $\eta(is)^{-24} = e^{2\pi s} + 24 + O(e^{-2\pi s})$. Once again we see that we encounter a tachyonic divergence in this ``closed-string'' channel (which would not appear in the superstring, but can be regulated here using analytic continuation).  In addition we  have a logarithmic divergence from the zero modes which is of the form
\beq
\int_1^{1/\ep_0^2}\frac{ds}{s} \sim \ln\,\left(\frac{2\pi \a'}{\ep^2}\right). 
\eeq
The massive modes do not give any divergences in this limit. This story entirely parallels the standard divergence structure of the one-loop (cylinder) diagram in open string theory. In that case, the above divergence coming from zero modes in the closed-string channel is cancelled out in the full unoriented, open plus closed string theory, provided we choose the right gauge group ($SO(32)$ in the superstring). We expect a similar cancellation to occur in the case of entanglement entropy, but this calculation will require a careful analysis of the entropy for closed strings.

In this section, we have demonstrated that in the free limit ($g_s \to 0$), the entanglement entropy computed using open string field theory is consistent with expectations from the effective field theory of the open-string excitations.  Namely, the answer we obtained was essentially the half-space entanglement entropy in a theory with an infinite number of free particles with the required masses and spins (minus degrees freedom removed by gauge symmetry). However, for finite $g_s$ and certainly in closed string theory, we expect this not to be generically true and the effective field theory approach should break down.\footnote{For instance, the one-loop diagram in closed string theory does not have UV divergences because of the restricted region of integration in the modular parameter $\tau$ of the torus, while a naive sum over the one-loop contributions from the particle excitations is UV divergent \cite{Polchinski:1985zf}. Similarly, we expect that in computing entanglement entropy for closed strings, one must carefully account for the correct region of integration for the modular parameter, something which is non-trivial in closed string field theory \cite{doi:10.1063/1.528569, zwiebach1991, Zwiebach:1992ie}.} For this reason, it is important to have an intrinsically stringy definition of the entanglement entropy, such as the one we have advocated using string field theory. %It would be very interesting, for instance, to compute the perturbative corrections to entanglement entropy coming from turning on a small value for the string coupling constant.

\section{Covariant Phase Space for Subregions} \label{CPS}
In the previous section, we neglected the gauge symmetry of open string field theory and the associated failure of factorization of the Hilbert space. In this section we will address this issue from the point of view of Witten's covariant open string field theory \cite{Witten:1985cc,Witten:1986qs} (see Appendix A for a lightning review). We will follow the method outlined by Donnelly and Freidel in \cite{Donnelly:2016auv} to analyze the phase space of string field theory on a subregion of the space of open string configurations. This will  lead us to an extended Hilbert space picture, with stringy edge modes localized at the entanglement cut. 

\subsection{Symplectic structure} \label{ps}
Witten's covariant open string field theory is constructed in terms of the string field 
$$\cA[X^{\mu}(\sigma), b_{\pm\pm}(\sigma), c_{\pm}(\sigma)]$$ 
which is an element of the one-string Hilbert space which we will call $\mathfrak{B}$ here.  Here $X^\mu$ are coordinates in target space and $b$ and $c$ are auxiliary ghost fields. It is possible to define a non-commutative star product $*$ on this space, together with the notion of ``integration'' $\int$, in terms of which the action takes the Chern-Simons form
\beq
S = \frac{1}{g_s^2} \int \left(\cA * Q\cA + \frac{2}{3} \cA*\cA*\cA\right),
\eeq 
where $Q$ is the BRST operator (Appendix A contains a brief review of the requisite background material, together with explicit expressions for $*$, $Q$ etc.). The equation of motion is given by
\beq\label{l1}
Q\cA + \cA*\cA = 0.
\eeq
Further, this action has the gauge symmetry
\beq\label{l2}
\delta \cA = Q\varepsilon +  \cA*\varepsilon - \varepsilon * \cA.
\eeq
Linearized about $\cA = 0$, equations \eqref{l1} and \eqref{l2} imply the statement that physical (on-shell) states are in the BRST cohomology. We will focus on this free limit in our discussion below. 

The important object for our discussion is the symplectic 2-form for OSFT constructed by Witten in \cite{Witten:1986qs} using the covariant phase space approach (see also \cite{SIOPSIS1987541}). In this formalism, we identify the space of solutions of the equations of motion as the \emph{phase space} $\mathcal{P}$; this is because the space of solutions is in one-to-one correspondence with the the space of initial conditions. In quantum field theory, since these initial conditions are functions of a Cauchy surface $\Sigma$ in spacetime, the symplectic 2-form can be written as an integral of a  density on $\Sigma$:
\beq
\bs{\omega} = \int_{\Sigma} J,\;\; \mathrm{with} \;\;\;d\star J = 0.
\eeq
The density $J$ must be conserved, because conservation ensures that the symplectic structure does not depend on the choice of $\Sigma$. Now we wish to construct such a symplectic 2-form for OSFT, which in the present case will be localized on a Cauchy surface $\cS \subset \cM_{\mathrm{open}}.$ For this purpose, let us first pick a Cauchy surface $\Sigma$ in the target spacetime, and let $\theta$ be defined as the following function of the string center of mass:
\beq \label{theta}
\theta(\Sigma) = \begin{cases} 1 & X_0^{\mu} > \Sigma \\ 0 & X_0^{\mu}  < \Sigma, \end{cases}
\eeq
with the notation $X_0^{\mu} > \Sigma$ ($X_0^{\mu} < \Sigma$) meaning that the point $X_0^{\mu}$ lies to the future (past) of $\Sigma$. Further, let us define $\bd $ to be the exterior derivative on the phase space $\mathcal{P}$, namely the space of solutions to the equation of motion $Q\cA + \cA*\cA = 0$. Then, for instance, $\bd \cA$ is a 1-form on the phase space, satisfying the linearized equation of motion around $\cA$. The symplectic 2-form on $\mathcal{P}$ is then given by
\beq 
\bs{\omega}= \int  \bd \cA* \left[Q, \theta(\Sigma) \right]\bd \cA,
\eeq
where we have left the wedge product between differential forms on $\mathcal{P}$ implicit. Firstly, note that $\bs{\omega}$ is a 2-form on $\mathcal{P}$. The BRST charge $Q$ has derivatives with respect to the string coordinates, and so the commutator $[Q, \theta(\Sigma) ]$ is proportional to a delta function for the center of mass lying on $\Sigma$; in other words the symplectic 2-form is localized on the Cauchy surface $\cS \subset \cM_{\co}$. Finally, $\bs{\omega}$ is independent of the choice of $\Sigma$  because given two surfaces $\Sigma_1$ and $\Sigma_2$:
\beq
\bs{\omega}_{\Sigma_1} - \bs{\omega}_{\Sigma_2} = \int Q\Big( \bd\cA *f(\Sigma_1,\Sigma_2)\bd\cA\Big) = 0,
\eeq
where the right hand side vanishes because the function $f(\Sigma_1,\Sigma_2)= \theta(\Sigma_1) - \theta(\Sigma_2) $ vanishes at infinity, and so $\int Q(\cdots)  = 0$. 

In \cite{Witten:1986qs} Witten used the string midpoint (instead of the center of mass) in the above construction. The choice of the midpoint is of course very natural in Witten's OSFT, but on the other hand from the point of view of the constituent fields, it seems more natural to use the center of mass of the string. We expect that at least in the free theory ($g_s \to 0$) which is the case we are focussing on, both the choices should be equivalent. At any rate, the symplectic form 
\beq 
\bs{\omega} = \int  \bd \cA* \mathfrak{D}_{0}\bd \cA = \left\langle \bd \cA | \mathfrak{D}_{0} \bd \cA\right\rangle,
\eeq
where $\mathfrak{D}_{0} = \left[ Q, \theta(\Sigma) \right]$ is defined in terms of the center of mass, has all the necessary properties, as explained above.\footnote{To see that this agrees with the standard symplectic structure for component fields, let us consider the photon field which appears at level one. Using \eqref{photon}, we can compute the symplectic 2-form:
$$\bs{\omega} = \frac{1}{2}\int  d^dx \,\delta(x^0)\Big(\bd A^{\mu}(\pa_0\bd A_{\mu})- (\pa_0\bd A_{\mu}) \bd A^{\mu} -2\bd A_0\bd f+2\bd f \bd A_0 \Big),$$
which upon using the equations of motion $2\bd f = \pa_{\mu} \bd A^{\mu}$ gives the standard symplectic form for the U(1) gauge field, up to boundary terms (at spatial infinity):
$$\bs{\omega} = \frac{1}{2}\int  d^dx \,\delta(x^0) \bd A^{\mu} \bd F_{0\mu} .$$} For simplicity, let us take the surface $\Sigma$ to be at $t_{\mu}X^{\mu}_0 = 0$, where we can take $t_{\mu} = (1,0,\cdots,0)$ or a more general time-like vector. Using equation \eqref{brstm}, we can compute the commutator $\mathfrak{D}_0$:
%\beqn \label{Qc}
%\left[ Q, \theta(\Sigma) \right] &=& \delta[x^0] \left(c^+\pa_+X^0 + c^-\pa_-X^0\right)(\pi/2)\nonumber\\
%&=& \frac{i}{\sqrt{2}}c^+(\pi/2)\delta[x^0]\left(-\frac{\delta}{\delta x^{0} }+ \pa_{\sigma}x_{0}\right)+ ic^-(\pi/2)\delta[x^0]\left(-\frac{\delta}{\delta x^{0} }- \pa_{\sigma}x_{0}\right). 
%\eeqn
\beq\label{Qc}
\mathfrak{D}_{0} =-\frac{i}{2}c_0\Big(t\cdot \a_0\,\delta(t_{\mu} X^{\mu}_0)+\delta(t_{\mu}X^{\mu}_0)\,t_{\mu} \a^{\mu}_0\Big)-i\delta(t_{\mu} X^{\mu}_0)\sum_{n\neq 0}c_{-n}\,t_{\mu} \a^{\mu}_n.
\eeq

To check for gauge invariance of the symplectic form at the level of the string fields, we must verify that it  vanishes when evaluated on a pure gauge mode
\beq
\bs{\omega}(Q \varepsilon, \Psi) =0,
\eeq
for any physical mode $\Psi$ satisfying $Q\Psi = 0$. This is the physical requirement that pure gauge modes are not dynamical. By explicit calculation, we obtain
\beqn
 \bs{\omega}(Q \varepsilon, \Psi) &=& \int \Big(   Q \varepsilon * \mathfrak{D}_{0}\Psi - \Psi * \mathfrak{D}_{0}Q\varepsilon \Big)\nonumber\\
&=&\int Q\Big(  \varepsilon *\mathfrak{D}_{0}\Psi - \Psi *\mathfrak{D}_{0}\ep \Big)\nonumber\\
%&=&\int Q\Big( \left[Q_A, \theta(\Sigma) \right] \left(\epsilon * \delta_2 \cA - \delta_1\cA * \epsilon-\epsilon * \delta_1 \cA + \delta_2\cA * \epsilon \right)\Big)\nonumber\\
&=& 0.
\eeqn
In the last line, we have used the fact that $ \int Q\chi= 0$, which is true as long as $\chi$ vanishes at infinity. We have therefore demonstrated that the pure gauge modes in the string field do not have any symplectic structure associated with them, and that $\bs{\omega}$ simply pulls-back to the moduli space of string fields modulo gauge transformations. This is generically true in the situation where $\Sigma$ has no boundaries, or the string field variations vanish at infinity. However, one has to be careful with such ``total derivative'' terms when we deal with the symplectic structure associated with {\it subregions}. Indeed, when we consider the phase space for subregions, this gives rise to non-trivial degrees of freedom at the boundary of the subregions.

\subsection{Symplectic structure for subregions}
Let us now consider the symplectic structure corresponding to a subregion $R \subset \Sigma$. For simplicity, we can take $R$ to be a half-space $X^1>0$. However, recall that by a subregion here we should mean a subregion in the Cauchy surface $\cS$ inside the space of open strings $\cM_{\co}$. As before, we pick the simplest extension\footnote{We are suppressing the dependence on ghosts here because the subregion includes the entire space of ghost configurations, i.e. we do not partition along the anti-commuting directions.}:
$$\cR = \left\{ X^{\mu}(\sigma) \in \cS \; | X^{1}_0 > 0 \right\}.$$
The symplectic two-form restricted to the region $\cR$ is then given by
\beq \label{subreg}
\bs{\omega}_\cR = \int  \bd \cA* \theta(\cR)\,\mathfrak{D}_0 \bd \cA  = \left\langle \bd \cA | \theta(\cR)\,\mathfrak{D}_0 \bd \cA\right\rangle,
\eeq
where by $\theta(\cR)$ we mean
\beq
\theta(\cR) = \begin{cases} 1 & X^{\mu}(\sigma) \in \cR \\ 0 & X^{\mu}(\sigma) \notin \cR.  \end{cases}
\eeq
\newcommand{\bep}{\boldsymbol{\delta}\varepsilon}
We now revisit the issue of  gauge-invariance. In order to be more systematic, let us choose coordinates $\cA = (a,\ep)$ on the space of string fields satisfying the equations of motion, where $a$ corresponds to the moduli-space of string fields modulo gauge transformations, and $\ep$ denotes the pure gauge directions:
\beq \label{para}
\cA = a +Q \ep.
\eeq
The one-form $\bd \cA$ then becomes
\beq
\bd \cA = \bd a + Q \bep,
\eeq
where we could think of $\bep$ as the linearized Maurer-Cartan form for the gauge group. Substituting this in equation \eqref{subreg}, we find
\beq
\bs{\omega}_\cR =  \int  \bd a* \theta(R)\,\mathfrak{D}_0 \bd a +\int \left( \bd a* \mathfrak{D}_1\mathfrak{D}_0 \bep - \bep *  \mathfrak{D}_1\mathfrak{D}_0\bd a -\bep*  \mathfrak{D}_1\mathfrak{D}_0\,Q\bep\right).
\eeq
The first term above is the standard symplectic form pulled back on to the moduli-space of string fields (modulo gauge transformations), and has support over the entire region $\cR$. The remaining terms indicate that the pure gauge modes do not entirely decouple in the presence of the entanglement cut. Note that the operator $\mathfrak{D}_1 = \left[Q, \theta(\cR) \right]$ is given by 
\beq\label{Qc1}
\mathfrak{D}_{1} =-\frac{i}{2}c_0\Big(r_{\mu} \a^{\mu}_0\,\delta(r_{\mu} X^{\mu}_0)+\delta(r_{\mu} X^{\mu}_0)\,r_{\mu} \a^{\mu}_0\Big)-i\sum_{n\neq 0}\delta(r_{\mu} X^{\mu}_0)c_{-n}\,r_{\mu} \a^{\mu}_n,
\eeq
where $r^{\mu}=(0,1,0,\cdots, 0)$ is the spacelike normal vector to the boundary $\pa R$. The delta functions in $\mathfrak{D}_1$ clearly imply that these gauge-dependent terms are localized on the boundary of the subregion.  In other words, the pure gauge modes become dynamical at the entanglement cut, giving rise to  \emph{stringy edge modes}. The fact that the symplectic 2-form for a subregion is not gauge-invariant is a standard feature of gauge theories  \cite{Donnelly:2011hn, Donnelly:2016auv}. It indicates that the Hilbert space does not factorize between subregions. Our discussion above shows that a similar situation arises in string field theory, and therefore care must be taken in defining the entropy of a subregion. The power of string field theory is that we need not revisit string quantization again in the presence of the entanglement cut to deduce these edge modes -- we can simply use gauge invariance in string field theory. 

One approach to deal with non-factorization of Hilbert spaces in gauge theories is the \emph{extended Hilbert space} formalism \cite{Buividovich:2008gq, Donnelly:2011hn, Casini:2013rba, Donnelly:2014fua, Radicevic:2014kqa, Ghosh:2015iwa, Donnelly:2016auv}. In this formalism, one embeds the physical Hilbert space as a subspace inside an extended Hilbert space:
\beq
\cH_{\mathrm{ext.}} = \cH_{\cR_b} \otimes \cH_{\mathrm{edge}} \otimes  \cH_{\overline{\mathrm{edge}}} \otimes \cH_{\overline{\cR_b}}.
\eeq
where $\cH_{\cR_b}$ and $\cH_{{\overline{\cR_b}}}$  are the Hilbert spaces in the bulk of the region $\cR$ and its complement. The extended Hilbert space includes two copies of edge modes (corresponding to $\pa \cR$ and $\pa \overline \cR$) and involves many non-physical states, but has the virtue that it factorizes naturally. Once the physical state of interest can be identified inside the extended Hilbert space, one can trace out the complement and define entanglement entropy in the standard way. 

In order to proceed, we need to understand how to construct $\cH_{\mathrm{edge}}$. To that end, it is useful to characterize the edge modes by making explicit the global symmetries acting on them. Note, for instance, that the symplectic form is invariant under the ``left-action''\footnote{One can also define a ``right-action'' 
$$\delta_R a = -Q \beta, \;\;\;\delta_R \ep = \beta$$
which is a mere redundancy of our parametrization \eqref{para}. The notation left and right is ambiguous in free string field theory, but becomes meaningful in the full non-linear version.}
 \beq
\delta_L a = 0, \;\;\;\delta_L \ep = -\alpha,
\eeq
because $\bs{\omega}_{\cR}$ depends only on $\bd \ep$ and not on $\ep$, where the former is invariant under this left-action because $\alpha$ is a constant with respect to the string phase space coordinates $(a,\varepsilon)$. We can easily write down the \emph{generators} of these boundary symmetries -- recall from Hamiltonian mechanics (written in terms of symplectic geometry) that this amounts to constructing charges $J[\a]$ such that 
\beq
 I_{V_{\a}} \bs{\omega}_R = \bd J[\a],
 \eeq
  where $I_X$ is the interior product on differential forms on $\mathcal{P}$ (given some vector field $X$ on $\mathcal{P}$). For instance, $I_{X} \bs{\omega} = \bs{\omega}(X, \cdot)$. The vector field $V_{\a}$ above generates the left action on phase space, and therefore satisfies $I_{V_{\a}} \bep = -\a.$ Now, a simple computation shows that
  \beq
I_{V_{\a}} \bs{\omega}_R = \int \left( \bd a* \mathfrak{D}_1\mathfrak{D}_0 \a + \a *  \mathfrak{D}_1\mathfrak{D}_0 \bd a +\a*  \mathfrak{D}_1\mathfrak{D}_0\,Q\bd \ep-\bd\ep*  \mathfrak{D}_1\mathfrak{D}_0\,Q\a\right),
\eeq
and so we can read off the currents from here
\beq
J[\a] = \int \left( a* \mathfrak{D}_1\mathfrak{D}_0 \a + \a *  \mathfrak{D}_1\mathfrak{D}_0 a +\a*  \mathfrak{D}_1\mathfrak{D}_0\,Q\ep-\ep*  \mathfrak{D}_1\mathfrak{D}_0\,Q\a\right).
\eeq
A further simple calculation shows that these currents satisfy the following current algebra:
%We can obtain the current algebra by computing $\delta_{\beta}J[\a]=I_{V_{\beta}}J[\a]$, which gives
\beq\label{ca}
 \left\{J[\a],J[\b]\right\}_{P.B}=\int \left(\a*  \mathfrak{D}_1\mathfrak{D}_0\,Q\b-\b*  \mathfrak{D}_1\mathfrak{D}_0\,Q\a\right).
\eeq
Therefore, the Hilbert space of the edge modes $\mathcal{H}_{\mathrm{edge}}$ should furnish a representation of this current algebra (in the $g_s \to 0$ limit). 

The next question is how to identify physical states (let's say the vacuum) inside the extended Hilbert space. The physical subspace of states is carved out by the ``quantum gluing'' condition:
\beq\label{qgc}
\left( J[\alpha] + \bar{J} [\alpha] \right) |\psi_{\mathrm{phys.}}\rangle = 0,
\eeq
where $J$ and $\bar{J}$ are the currents corresponding to the boundary of $\cR$ and $\bar{\cR}$ respectively. The quantum gluing condition is simply the statement that the global symmetries acting of the edge modes should not be visible in physical states. However, in bosonic string theory (similar to the situation in Maxwell theory \cite{Donnelly:2014fua}), this condition has many solutions -- we can think of the various solutions of the quantum gluing condition as corresponding to boundary conditions at the entanglement cut. So one has to give some further dynamical input in order to identify the required physical state in the extended Hilbert space, thus complicating the calculation considerably. More precisely, any physical state corresponds to a probability distribution over the boundary conditions. For example, in the case of the vacuum state in Maxwell theory, the probability distribution can be fixed by appealing to Euclidean path-integral arguments \cite{Donnelly:2014fua}. In order to repeat a similar analysis in string field theory would require a careful treatment of Euclidean path-integral methods in OSFT. It would be interesting to try to push this calculation to completion, because we expect that the contribution to the entanglement entropy coming from edge modes should reproduce the contact terms typically found in computations of the conical entropy \cite{Kabat:1995eq, He:2014gva}. In the case of closed string theory, these contributions might even be related to the gravitational Bekenstein-Hawking entropy. In this case, it is an interesting question to relate the edge modes we have discussed above to the open strings on the horizon which appeared in the work of Susskind and Uglum \cite{Susskind:1994sm}. In the rest of this paper, we will merely demonstrate how the above computation of entropy using the extended Hilbert space works in a simpler example, instead of bosonic strings, leaving the latter case for future work.

The simple model in question, which fits within the algebraic framework of Witten's string field theory described above, is ordinary $U(1)$ Chern-Simons theory. In this case, we replace $\mathfrak{B}$ with the space of differential forms on some 3-manifold $X$, the BRST charge $Q$ with the exterior derivative $d$, the product $*$ with the wedge product on differential forms, and $\int$ with $\frac{k}{4\pi}$ times integration on $X$, with $k$ some positive integer. Indeed, this is no accident -- Chern-Simons theory on $S^3$ can be realized as the effective field theory of zero-modes on the worldvolume of A-branes in the A-type topological string theory on $T^*S^3$ \cite{Witten:1992fb} (and under some conditions outlined in \cite{Witten:1992fb}, this description is exact). Importantly, the above extended Hilbert space formalism can be pushed to its logical conclusion in this simple toy model, as we will now review following \cite{Fliss:2017wop} (see also \cite{Geiller:2017xad, Wong:2017pdm}). For simplicity, let us consider the spatial slice $\Sigma$ to be a 2-sphere $S^2$, and let the subregion $R$ be a disc. In this case, the current algebra in equation \eqref{ca} becomes (upon making the replacement $\{,\}_{P.B} \to i\left [,\right]$)
\beq
\Big [J[\alpha] , J[\beta] \Big]= \frac{ik}{4\pi} \oint_{\pa R} d\theta \left(\a  \pa_{\theta}\b-\b  \pa_{\theta}\a\right),
\eeq
where the above integral is only over the boundary of the subregion (because $\mathfrak{D}_0 = \left[Q, \theta(x^0)\right] = \delta(x^0) dx^0$ and similarly for the spacelike direction orthogonal to the cut). Switching to momentum modes on the circle, we identify this as the $U(1)_k$ chiral Wess-Zumino-Witten current algebra:
\beq
\left[J_m, J_n\right] = \frac{k}{2} n\,\delta_{n+m,0}.
\eeq
The quantum-gluing condition \eqref{qgc}, $\left(J_n + \overline{J}_{-n} \right) |\psi_{\mathrm{phys.}}\rangle = 0$, is solved by the \emph{Ishibashi} states:
\beq
| q \rangle\rangle  = \sum_{N=0}^{\infty} \sum_{z\in \mathbb{Z}} \sum_{\ell=1}^{d_N} | q, z, N, \ell \rangle \otimes |\overline{q, z, N, \ell}\rangle,
\eeq
where $q = 0, 1, 2 \cdots, k-1$ is the charge corresponding to an integrable representation, and $z,N,\ell$ label the descendents. The choice of $q$ determines which state we are considering in the Chern-Simons theory. For $q=0$, we obtain the vacuum state. For non-trivial $q$, we obtain a state with Wilson lines with charges $q$ and $-q$ piercing through $R$ and $\overline{R}$ respectively. We can now compute a well-defined  entanglement entropy between $R$ and $\bar{R}$. This is essentially the computation of left-right entanglement entropy in Ishibashi states carried out in \cite{2012PhRvL.108s6402Q, PandoZayas:2014wsa, Das:2015oha} (see also \cite{Schnitzer:2015gpa, Zayas:2016drv, Vancea:2016tkt, Vancea:2017pom}). These papers showed by explicit computation that the left-right entanglement entropy in the Ishibashi state $| q  \rangle\rangle$ exactly reproduces the topological entanglement entropy of the corresponding state in Chern-Simons theory on $S^2$ bi-partitioned into two discs (which has also previously been computed by other methods in \cite{Kitaev:2005dm, Levin:2004mi, Dong:2008ft}).  

\section{Discussion}

In this paper, we used string field theory to define and compute entanglement entropy between spatial subregions of the target space in open string theory.   We demonstrated that in the free limit, this entropy is a sum over the one-loop entropies of particle excitations of the string, and that the inclusion of the tower of stringy degrees of freedom softens the divergences in the entanglement entropy.   Finally, we adapted the formalism of Donnelly and Freidel \cite{Donnelly:2016auv} who studied entanglement entropy in gauge field theories to study the covariant phase space of subregions in the string theoory target space  and demonstrated the existence of novel stringy edge modes.  We argued that these edge modes will contribute to entanglement entropy in string theory.   

\begin{figure}[!h]
\centering
\includegraphics[height=5cm]{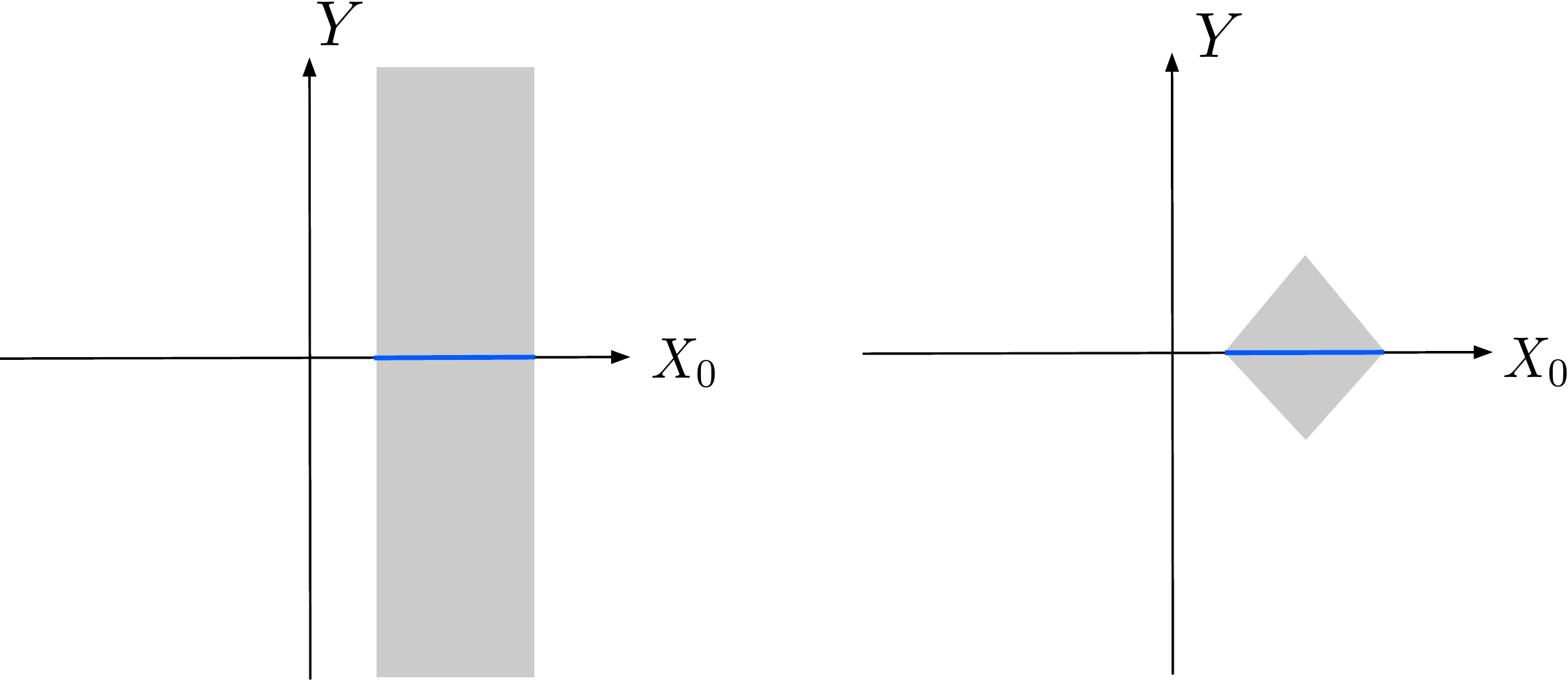}
\caption{\small{Consider the simplified setup where we discreteize the string to two points $X^{\mu} = (x^{\mu}_1, x^{\mu}_2)$. In terms of the center of mass $X_0 = \frac{1}{2}(x_1+x_2)$ and the separation $Y = \frac{1}{2}(x_1- x_2)$, we can extend the subregion (in blue) from target space to the entire configuration space in multiple ways. On the left, we count all strings with center of mass in the blue region, while on the right we count all strings which are entirely in the blue region.  Note that while the subregion on the left looks non-compact, the string tension implies that the wavefunction of physical states of the string will be exponentially localized around $Y=0$. }}
\end{figure}

In order to compute the entropy in OSFT, we made a choice for how to extend the entanglement region from the target spacetime to the full configuration space of open strings $\cM_{\mathrm{open}}$ -- namely, we included all strings with their center of mass inside the subregion. This extension is not unique, and  the specific result of the entanglent entropy computation will depend of the choice of extension. For instance, we could have counted all strings which lie {\it entirely} within the subregion (see figure 3). However, we expect that independent of how we extend the subregion, string theory will soften the UV divergences, and that stringy edge modes should always exist because of the gauge symmetry of string field theory. It is desirable to better understand the dependence of the entanglement entropy on the choice of extension. It might also be interesting to understand the connection of ideas presented in this paper with the studies of causality in string field theory \cite{Martinec:1993jq, Lowe:1993ps, Lowe:1994ns, Lowe:1995ac, Hata:1996hd, Erler:2004hv}. Another question of interest is the finite $g_s$ corrections -- entanglement entropy in interacting, non-local theories potentially has an interesting cross-over behavior between area law and volume law depending on whether the subregion size is larger or small compared to the scale of non-locality \cite{Fischler:2013gsa, Karczmarek:2013xxa, Rabideau:2015via, Chen:2017kfj}. It would be interesting to check whether open string theory exhibits this phenomenon, perhaps perturbatively in $g_s$.

The general form of our calculations suggests that in a full open + closed super-SFT with the correct gauge group the entanglement entropy may be strictly finite.  If true that would be very interesting and useful. This may be related generally to the idea of the holographic principle and the idea that the number of degrees of freedom in quantum gravity in a finite region should actually be finite.   In particular if we consider a black hole, it would be very interesting to understand the entanglement entropy in string theory across the horizon.  This was the focus of early attempts to understand the origin of black hole thermodynamics in quantum gravity, but those efforts were carried out in first-quantized string theory using the replica method and were challenged by the presence of off-shell conical backgrounds that may not be well-defined \cite{Susskind:1994sm}.  Here we have outlined an approach, albeit in open string field theory, that may allow a  precise calculation of entanglement across a horizon that does not face these challenges. For this purpose however, we need to extend this approach to closed string theory. An important subtlety in this case is that the definition of a subregion in a gravitational theory requires an asymptotic boundary or a horizon, since only diffeomorphism invariant constructs are meaningful. We expect therefore that the choice of a subregion in loop space should involve a careful, diffeomorphism invariant construction. Secondly, it is also presently unclear how closed string field theory regulates the ultraviolet divergences in entanglement entropy. Finally, at a more abstract level, we could take the point of view that a consistent string theory is a 2d CFT (with special properties) and that spacetime is a meaningful concept only in certain special limits. It would be interesting to try to reconcile some of these points with the ideas presented in this paper. 

\subsection*{Acknowledgements}
We would like to thank Pawel Caputa, Falk Hassler, Jonathan Heckman, Arjun Kar, Rob Leigh, Aitor Lewkowycz, Djordje Minic, Charles Rabideau and Tadashi Takayanagi for helpful discussions. Research funded by the Simons Foundation (\#385592, VB) through the It From Qubit Simons Collaboration, and the US Department of Energy contract \#FG02-05ER-41367. 

\appendix
\section{Basic Review of Covariant String Field Theory}
In this appendix, we will briefly review Witten's covariant open string field theory, following \cite{Witten:1985cc} (see also \cite{GROSS19871,Taylor:2003gn, Taylor:2006ye} and references there-in). This is entirely standard material, and is included here only for completeness. The worldsheet action is given by
\beq
S_X = \frac{1}{4\pi} \int d^2\sigma\; \pa_{\a}X^{\mu}(\tau,\sigma) \pa^{\a} X_{\mu}(\tau,\sigma),
\eeq
%The mode expansion at $\tau = 0$ is given by
%\beq
%X^{\mu}(\sigma)  = X^{\mu}_0 + \sum_{n=1}^{\infty} \frac{1}{\sqrt{n}} \cos(n\sigma)\,X^{\mu}_n
%\eeq
while the ghost action (arising from gauge-fixing the $\mathrm{Diff} \times \mathrm{Weyl}$ gauge symmetries on the worldsheet) is given by
\beq
S_{gh} = \frac{1}{\pi} \int d^2\sigma\; \left(c_-\pa_- b_{++} + c_+ \pa_+ b_{--}\right)
\eeq
where $\pa_{\pm} = \frac{1}{\sqrt{2}} \left(\pa_{\tau} \pm i \pa_{\sigma}\right)$. Open string boundary conditions are $c_+ = c_-$ and $b_{++} = b_{--}$ at the end points $\sigma = 0, \pi $. With these boundary conditions, both $b$ and $c$ have precisely one zero mode. %The stress tensors are 
%\beq
%T^{X}_{++} = \pa_{+}X^{\mu} \pa_{+}X_{\mu},\;\;T^{gh}_{++} = c_- \pa_+b_{++} + 2(\pa_+ c_-)b_{++}
%\eeq
%Similarly, $T_{--}$ is given by $+ \leftrightarrow -$, and $T_{+-} = 0$. 
The BRST current can be written in terms of the worldsheet Virasoro generators as:
%\beq
%J_{\pm} = c_{\mp} \left(T^X_{\pm \pm} + \frac{1}{2} T^{gh}_{\pm \pm}\right),
%\eeq
%while the BRST charge is given by
%\beq
%Q = \int_0^{\pi} d\sigma\;J_0(\sigma).
%\eeq
%For concreteness, let us write out some of the terms in $Q$ which will be relevant later on:
%\beqn \label{brst}
%Q &=& \int_0^{\pi} d\sigma\,\left(c_-T_{++} + c_+ T_{--} \right) \\
%&=& \sum_{a= +, -} \int_0^{\pi} d\sigma\, c_{a} \left(- \frac{\delta^2}{\delta X^{\mu}(\sigma)\delta X_{\mu}(\sigma)} -a \left\{ \frac{\delta}{\delta X^{\mu}(\sigma)}, \pa_{\sigma}X^{\mu}\right\} - \pa_{\sigma}X^{\mu}\pa_{\sigma}X_{\mu}\right) +\mathrm{ghosts\;terms}\nonumber
%\eeqn
%t is also convenient to rewrite $Q$ in terms of the worldsheet Virasoro generators:
\beq\label{brstm}
Q = c_0(L_0 -1 ) + \sum_{n\neq 0} c_{-n}L_n - \frac{1}{2}\sum_{n,m}(n-m):c_{-n}c_{-m}b_{n+m}:
\eeq
where recall that the worldsheet-Virasoro generators can be written in terms of the string-oscillator modes as%\footnote{Here $\alpha_0^{\mu} = p_0^{\mu}$.} 
\beq
L_n = \frac{1}{2}\sum_{m=-\infty}^{\infty} : \a_{n-m} \a_m:.
\eeq
Note also that following standard convention, we have used the ``folding trick'' to rewrite the two copies of ghosts (corresponding to $\pm$) in terms of a single copy of the chiral closed string modes and then Fourier expanded these in terms of $c_n$ and $b_n$.

When $D=26$, the BRST charge is nilpotent, i.e. $Q^2 = 0$. The quantization of a single string in the BRST formalism is accomplished by requiring the physical Hilbert space to be in the BRST cohomology
\beq
Q |\psi\rangle = 0 ,
\eeq
with two states $\psi$ and $\psi' $ considered equivalent if $|\psi'\rangle = |\psi\rangle + Q |\epsilon\rangle$. Additionally, we also require physical states to be annihilated by all ghost and anti-ghost annihilation operators $b_{n}, c_{m} \; \forall \, n,m>0$, and also by $b_0$. Following standard notation, we will denote the physical vacuum in the ghost zero-mode sector by $| \downarrow \rangle$: 
\beq
b_0 |\downarrow \rangle = 0.
\eeq 
and assign it ghost number $-1/2$. %Acting on such states, the BRST operator reduces to
%\beq
%Q = c_0(L_0 -1 ) + \sum_{n>0} c_{-n}L_n,
%\eeq
%and therefore the BRST closed condition reduces to the standard Virasoro constraints. 

%In fact, it is more convenient to use a Bosonized description for the ghosts, where they are described by a single worldsheet scalar $\phi$:
%\beq
%S= \frac{1}{2\pi} \int d^2\sigma\;\left(\pa_{\a}\phi  \pa^{\a}\phi - 3i R\, \phi\right)
%\eeq
%The mode expansion for $\phi$ at $\tau= 0$ is given by
%\beq
%\phi(\sigma) = \phi_0 + \sum_{n>0} \sqrt{\frac{2}{n}} \cos(n\sigma)\,\phi_n
%\eeq
%We can also decompose this as
%\beq
%\phi_{\pm} (\sigma) = \phi_0 \pm \sigma\left(p_0 + 1/2\right) + \sum_{n>0} \sqrt{\frac{2}{n}}\Big( \cos(n\sigma) \phi_n \pm \sin(n\sigma) \pi_n \Big)
%\eeq
%In terms of these fields, the ghosts are given by
%\beq
%c(\sigma) = :e^{i\phi_+(\sigma)} :,\;\; b(\sigma) = :e^{i \phi_-(\sigma)}:
%\eeq

\begin{figure}[t]
\centering
\includegraphics[height=3.5cm]{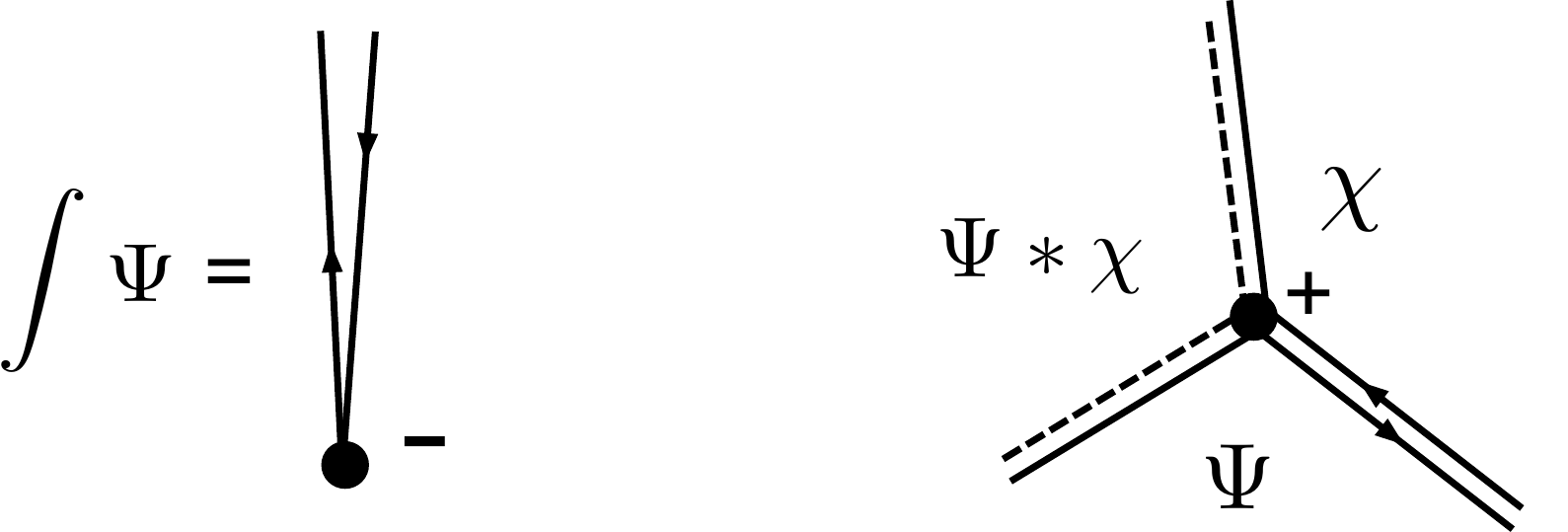}
\caption{\small{\textsf{Pictorial representations of the operations $\int$ and $*$. The black dot with $\pm$ signs indicates an insertion of $e^{\pm  \frac{3i}{2}\phi}$ at the midpoint. }} \label{alg}}
\end{figure}

Witten's formulation of open-string field theory (OSFT) is in terms of a \emph{non-commutative gauge theory}. It is convenient to bosonize the ghosts $(b,c) \to \phi$ in order to introduce this version of OSFT, but for our purposes the details of the ghosts will not be very important, so we do not give all the details of the bosonization map here (see \cite{Witten:1985cc} for the relevant details). The gauge field of this non-commutative gauge theory lives in an algebra $\mathfrak{B}$, where the elements of $\mathfrak{B}$ are states of  the one-string Hilbert space. Equivalently, one can think of $\mathfrak{B}$ as consisting of string fields 
\beq 
 \Psi[X^{\mu}(\sigma), \phi(\sigma)] = \left\langle X^{\mu}(\sigma), \phi(\sigma) | \Psi \right\rangle
 \eeq
on $\cM_{\co}$ -- the space of open-string configurations on target space (plus ghost configurations).  Recall that Fourier modes of the string field  create and annihilate single string states in the OSFT Hilbert space.  The algebra $\mathfrak{B}$ has a $\mathbb{Z}$-grading
$$\mathfrak{B} = \oplus_{n =-\infty}^{\infty} \mathfrak{B}_n,$$ 
where  $n=G+3/2$ with $G$ being the ghost number. As explained before, physical states have ghost number $G=-1/2$ and therefore live in $\mathfrak{B}_1$; consequently we may refer to them as \emph{one-forms} (more generally $n$-forms are elements of $\mathfrak{B}_n$). We now describe two operations which will be central to the construction of the OSFT:
\begin{itemize}
\item Integration, $\int : \mathfrak{B} \to \bC$, and
\item Product, $*: \mathfrak{B} \times \mathfrak{B} \to \mathfrak{B}$. 
\end{itemize}
The integration is defined as
\beq
\int \Psi  = \int DX^{\mu}(\sigma) D\phi(\sigma)\,e^{-\frac{3i}{2}\phi(\frac{\pi}{2})} \prod_{\sigma=0}^{\pi/2}\delta [X^{\mu}(\sigma) - X^{\mu}(\pi - \sigma)]  \prod_{\sigma=0}^{\pi/2}\delta [\phi (\sigma) - \phi (\pi - \sigma)]  \Psi[X^{\mu}(\sigma), \phi(\sigma)],
\eeq
namely one simply sews together the two halfs of the string with an insertion of the operator $e^{-\frac{3i}{2}\phi(\frac{\pi}{2})}$ at the midpoint (see figure \ref{alg}). %We can also rewrite this expression in terms of the oscillators as
%\beq
%\int \Psi  = \int DX^{\mu}_n D\phi_n\,e^{-\frac{3i}{2}\left(\phi_0+ \sum_{k>0,\,k\,\mathrm{even}}\sqrt{\frac{2}{k}}(-1)^{k/2}\phi_k\right)} \prod_{m>0,m\;\mathrm{odd}}\delta [X^{\mu}_m] \delta [\phi_m] \; \Psi[X^{\mu}_n, \phi_n].
%\eeq
The star product is defined by
\beqn
(\Psi*\chi)[X^{\mu}(\sigma), \phi(\sigma) ] &=& \int \prod_{i=1}^2DX_iD\phi_i\,e^{+\frac{3i}{2}\phi(\pi/2)} \Psi[X_1^{\mu},\phi_1] \chi[X_2^{\mu}, \phi_2]\\
&\times & \prod_{\sigma=0}^{\pi/2}\delta[X^{\mu}(\sigma) - X_1^{\mu}(\sigma)]\delta[X_1^{\mu}(\pi - \sigma) - X_2^{\mu}(\sigma)]\delta[X_2^{\mu}(\pi - \sigma) - X^{\mu}(\pi-\sigma)]\nonumber\\
&\times & \prod_{\sigma=0}^{\pi/2}\delta[\phi(\sigma) - \phi_1(\sigma)]\delta[\phi_1(\pi - \sigma) - \phi_2(\sigma)]\delta[\phi_2(\pi - \sigma) - \phi(\pi- \sigma)],\nonumber
\eeqn
or in other words we sew the right half of the first string with the left half of the second string, this time inserting $e^{+\frac{3i}{2}\phi}$ at the midpoint (see figure \ref{alg}).\footnote{The operator insertions at the midpoint are important to ensure ghost-number conservation.} 
Some important properties of the non-commutative algebraic structure we have encountered above are listed here:
\beq
Q(A*B) = (QA)*B + (-1)^{n_A}A*(QB),
\eeq
\beq
A*(B*C) = (A*B)*C,
\eeq
\beq
\int A*B = (-1)^{n_An_B} \int B*A,
\eeq
for $A, B, C \in \mathfrak{B}$.  Also, explicit calculation shows that 
\beq
\int Q f = 0
\eeq
for any $f$ that vanishes at infinity.   It is also worth noting that the operation $\int \Psi * \chi$ is quite simple:
\beq
\int \Psi*\chi = \int [DX D\phi]\, \Psi[X(\pi - \sigma), \phi(\pi - \sigma)] \chi[X( \sigma), \phi( \sigma)].
\eeq
If we further impose the reality condition 
\beq
\Psi[X^{\mu}(\pi-\sigma),\phi(\pi-\sigma)] = \Psi^*[X^{\mu}(\sigma),\phi(\sigma)]
\eeq
on the string fields, then it becomes clear that
\beq
\int \Psi*\chi = \langle \Psi |\chi \rangle.
\eeq

Having defined the algebra $\mathfrak{B}$ and its attendant operations, Witten wrote down the following action for open string field theory in terms of the string gauge field $\cA \in \mathfrak{B}_1$:
\beq 
S_{OSFT}[\cA] = \frac{1}{g_s^2}\int \left( \cA*Q\cA + \frac{2}{3} \cA*\cA*\cA \right).
\eeq
The equation of motion corresponding to this action is given by
\beq\label{e1}
Q \cA + \cA * \cA = 0.
\eeq
String interactions appear in the second term.
Further, the action is naturally gauge invariant under the infinitesimal gauge transformation
\beq \label{e2}
\delta_{\varepsilon} \cA = Q \varepsilon + \cA*\varepsilon - \varepsilon *\cA ,
\eeq
where $\varepsilon \in \mathfrak{B}_0$ is a zero form (i.e. has ghost number -3/2). Equations \eqref{e1} and \eqref{e2} linearized around $\cA=0$ give the BRST cohomology conditions on physical states. 

It is helpful to see how the above equation of motion and gauge symmetry lead to the Maxwell equation and $U(1)$ gauge symmetry in terms of the string-oscillators at level one. We will work with the free string field, which satisfies $Q\cA = 0$, and only keep track of the photon in our string field.  This amounts to expand the string field up to the first oscillator level and ignoring the tachyon (the zeroth oscillator level):
\beq \label{photon}
|\cA \rangle = \int \frac{d^{D}k}{(2\pi)^D} \left(-iA_{\mu}(k) \alpha^{\mu}_{-1} + f(k) c_0b_{-1}\right)|0,k, \downarrow \rangle ,
\eeq
where $A_{\mu}$ corresponds to the photon and $f$ is an auxiliary field (whose origin will become clear below). Using equation \eqref{brstm}, the equation of motion is given by
\beq
Q |\cA \rangle = \int \frac{d^{D}k}{(2\pi)^D}\Big( \frac{1}{2}\left(-ik^2 A_{\mu}+ 2k_{\mu}f \right) \a^{\mu}_{-1} c_0 -\left(i k_{\mu}A^{\mu} -2f\right)c_{-1}\Big) |0, k,\downarrow \rangle = 0.
\eeq
The equation of motion therefore imposes 
$$\Box A_{\mu} = 2\pa_{\mu}f,\;\;2f = \pa_{\mu}A^{\mu},$$
which together of course constitute Maxwell's equations. The $U(1)$ gauge transformation comes from the following OSFT gauge transformation:
\beq
|\cA \rangle \to |\cA\rangle + Q |\varepsilon\rangle,\;\;\; |\varepsilon\rangle = \int \frac{d^Dk}{(2\pi)^D} \varepsilon(k) b_{-1}|0, k, \downarrow \rangle,
\eeq
More explicitly, we find
\beq
 Q |\varepsilon\rangle = \int \frac{d^{D}k}{(2\pi)^D} \varepsilon(k) \left(\frac{1}{2}k^2 c_0 b_{-1}+k_{\mu} \alpha^{\mu}_{-1}\right)|0,k\rangle,
\eeq
which of course corresponds to shifting 
$$A_{\mu} \to A_{\mu} +\pa_{\mu}\varepsilon,$$
as expected. Note that in addition, it also shifts $f \to f+ \frac{1}{2}\Box \varepsilon$. It is clear from the above discussion, that the field $f$ is an auxiliary field which upon integrating out leads to the covariant Maxwell theory. We can think of ``fixing'' $f$ as a choice of gauge-fixing. For example, fixing $f= 0$ is equivalent to the Lorentz gauge. 
 
%We can also write down the finite version of the gauge transformation. Let us introduce the identity operator 
%\beq
%\bI[X^{\mu}(\sigma),\phi(\sigma) ]  = e^{-\frac{3i}{2}\phi(\pi/2)}\prod_{\sigma=0}^{\pi/2}\delta[ X^{\mu}(\sigma) - X^{\mu}(\pi - \sigma)]\delta[ \phi(\sigma) - \phi(\pi - \sigma)]
%\eeq
%which has the property
%\beq
%\Psi * \bI = \bI * \Psi = \Psi.
%\eeq
%for any $\Psi$, and by virtue of the ghost insertion at the midpoint, has the ghost number $-3/2$ (i.e. $d_{\bI} = 0$). Now we define the finite gauge transformation
%\beq
%U = \exp_*(-\varepsilon) \equiv \bI - \varepsilon + \frac{1}{2!} \varepsilon * \varepsilon + \cdots,
%\eeq
%\beq
%\cA_U = U* \cA * U^{-1} - QU * U^{-1}. 
%\eeq
%It is straightforward to check that under a gauge transformation, the action $S_{OSFT}[\cA]$ transforms up to a total derivative of the form $Q(\chi)$ which vanishes upon integration, and is therefore gauge invariant. 

%\bibliographystyle{uiuchept}
%\bibliography{SFT}
\providecommand{\href}[2]{#2}\begingroup\raggedright\endgroup

\end{document}